\documentclass[12pt]{iopart}
\usepackage{graphicx}
\usepackage{dcolumn}
\usepackage{bm}

\def\>{\rangle}
\def\<{\langle}

\begin{document}

\title{A heralded two-qutrit entangled state}

\author{Jaewoo Joo}
\address{Institute for Quantum Information Science,
University of Calgary, Alberta T2N 1N4, Canada}
\address{Blackett Laboratory, Imperial College London, Prince
Consort Road, London, SW7 2BW, UK}

\author{Terry Rudolph}
\address{Blackett Laboratory, Imperial College London, Prince
Consort Road, London, SW7 2BW, UK} \address{Institute for
Mathematical Sciences, Imperial College London, London, SW7 2BW,
UK}

\author{Barry C. Sanders}
\address{Institute for Quantum Information Science,
University of Calgary, Alberta T2N 1N4, Canada}

\date{\today}

\begin{abstract}
We propose a scheme for building a heralded two-qutrit entangled
state from polarized photons. An optical circuit is presented to
build the maximally entangled two-qutrit state from two heralded
Bell pairs and ideal threshold
detectors. Several schemes are discussed for constructing the two
Bell pairs. We also show how one can produce an unbalanced
two-qutrit state that could be of  general purpose use in
some protocols. In terms of applications of the maximally
entangled qutrit state, we mainly focus on how to use the state
to demonstrate a violation of the
Collins-Gisin-Linden-Massar-Popescu inequality under the restriction of measurements which can be performed using linear optical elements and photon counting. Other possible
applications of the state, such as for higher dimensional quantum
cryptography, teleportation, and generation of heralded two-qudit
states are also briefly discussed.
\end{abstract}

 \maketitle
\section{Introduction}

Sources of entangled photons are important ingredients for
optical quantum information processing. A convenient way to
generate maximally entangled photons is based on spontaneous
parametric down-conversion (PDC) that has been used to demonstrate
prototypes of many applications in quantum information processing
\cite{Nielsen,RMPKok}. However, to satisfy scalability
requirements for quantum computing and other protocols,
deterministic or at least ``heralded'' (event ready) photon
sources are required \cite{Kok00,HeraldEBP}.

The basic unit of quantum information is the qubit. However a
variety of results \cite{advantage,advantage2,Peres,Eisenberg}
have shown that multi-level systems - qu$d$its ($d$; dimensions)
have advantages over qubits in certain circumstances. In
particular, it is experimentally feasible to control several
photons very precisely as such higher dimensional objects
\cite{Andrew,Lanyon,Bogdanov}, and photonic qutrits have been
implemented using PDC in orbital angular momentum, polarization,
multi-mode, and energy-time
\cite{Vaziri02,Howell,Zeilinger97,Gisin04}.

We analyze a scheme for building a heralded two-qutrit state from
polarized photons. To achieve the qutrit state, generation of two
heralded Bell pairs and four quantum memories are required. The
Bell pairs are mixed at two beam-splitters (BSs), and two null
detections herald the desired qutrit state in two spatial modes.
The detector is assumed an ideal threshold detector, which does
not need to be number resolving - it need only trigger on
non-absorption versus absorption. We discuss how one can produce
an unbalanced two-qutrit entangled state that may be of more
general purpose use in some protocols \cite{advantage}. In terms
of applications of the qutrit state, we focus on how to use it to
demonstrate a violation of the
Collins-Gisin-Linden-Massar-Popescu (CGLMP) inequality
\cite{Bellinequal02}.

The paper is organized as follows. In Section \ref{sec2}, we
introduce notation and also discuss how a typical state generated
by Type-II PDC includes a two-qutrit state as a third-order
process. In Section \ref{sec3}, we discuss several schemes to
construct two heralded Bell pairs and show an optical circuit to
build a heralded two-qutrit state from the Bell pairs. In Section
\ref{sec4} we discuss the ability of the single-qutrit state to
violate the CGLMP inequality in the case that the measurements
performed are done with linear optics (which restricts the
possible qutrit operations from $U(3)$ to those of a
3-dimensional representation of $SU(2)$). Other possible
applications of the qutrit state, such as for higher dimensional
quantum cryptography, teleportation, and generation of heralded
two-qudit states are also briefly discussed. In Section
\ref{sec5}, we conclude by summarizing the results.

\section{Background}
\label{sec2} In this section, we present our notation for
maximally entangled states in polarization and a typical
generation scheme of polarization entangled states using the PDC
method that has been used for many important proof-of-principle
demonstrations of quantum information processing. We show how
from Type-II PDC photons, a two-qutrit entangled state in
polarization can be probabilistically observed by post-selection.

\subsection{Maximally entangled states in polarization}
Let us first define a qudit state with basis $|i\rangle$
($i=0,1,...,d-1$). A maximally entangled bipartite state between
$A$ and $B$ for qudits is represented by
\begin{equation}
\label{eq:Dstate01} |\psi^{d} \rangle_{ AB} = {1 \over \sqrt{d}}
\sum^{d-1}_{i=0} | i \rangle_A | i \rangle_B \in {\cal H}_d
\otimes {\cal H}_d = d \otimes d.
\end{equation}
$| i \rangle$ is an orthonormal basis of ${\cal H}_d$. One of the
Bell pairs is represented by
\begin{equation} |\psi^{2}\rangle_{
AB} = {1\over\sqrt{2}} (| 0 \rangle_A | 0 \rangle_B + | 1
\rangle_A | 1 \rangle_B ),
\end{equation}
and
\begin{equation}
|\psi^{3}\rangle_{ AB} = {1\over\sqrt{3}} (| 0 \rangle_A | 0
\rangle_B + | 1 \rangle_A | 1 \rangle_B + | 2 \rangle_A | 2
\rangle_B ).
\end{equation}
We denote a photonic state
\begin{equation}
\label{eq:photonicpolarizationstate} |m,
n\rangle_A={(a^{\dagger}_H)^m (a^{\dagger}_V)^n \over
\sqrt{m!n!}} | 0 , 0 \rangle_A
\end{equation}
consisting of $m$ horizontal photons and $n$ vertical photons in
spatial mode $A$ ($a^{\dagger}_{H (V)}$; creation operator for
horizontal (vertical) polarization, $| 0 , 0 \rangle$; a vacuum
state), the photonic state $|m, n\rangle_A$ spans ${\cal H}_d$ as
an orthonormal basis. Thus, the state in Eq.~(\ref{eq:Dstate01})
is equal to
\begin{equation}
\label{eq:photonDstate01} |\psi^{d} \rangle_{ AB} = {1 \over
\sqrt{d}} \sum^{d-1}_{i=0} | d-1-i ,i \rangle_A | d-1-i ,i
\rangle_B.
\end{equation}
For example, the Bell state is represented by
\begin{equation}
\label{Bellpair} |\psi^{2}\rangle_{ AB} = {1\over\sqrt{2}} (| 1 ,
0 \rangle_A | 1 , 0 \rangle_B + | 0 , 1 \rangle_A | 0 , 1
\rangle_B ).
\end{equation}
A maximally entangled state for qutrits is given by
\begin{equation}
\label{eq:newstate03} |\psi^{3}\rangle_{ AB} = {1\over\sqrt{3}} (|
2 , 0  \rangle_A | 2 , 0 \rangle_B + | 1 , 1 \rangle_A | 1 , 1
\rangle_B + | 0 , 2 \rangle_A | 0 , 2 \rangle_B ).
\end{equation}
Therefore, three basis vectors for qutrits can be defined by
$|0\rangle \equiv |2 , 0 \rangle $, $|1\rangle \equiv |1 , 1
\rangle $, and $|2\rangle \equiv |0 , 2 \rangle$.

\subsection{Photon-pair source in PDC}

Using Type-II PDC scheme, the superposition of the maximally
entangled states can be generated. A PDC photonic state in modes
$A$ and $B$ is equal to
\begin{equation}
\label{SPDCphotons} |\Psi^{\rm PDC} \rangle_{AB} = {\rm e}^{-{\rm
i} H t} |0,0\rangle = {1 \over \cosh^2 \tau} \sum^{\infty}_{d=1}
\sqrt{d} \tanh^{d-1} \tau | \psi^{d}_{-} \rangle_{A B},
\end{equation}
where
\begin{equation}
H={\rm i} \kappa(a^{\dagger}_H b^{\dagger}_V-a^{\dagger}_V
b^{\dagger}_H) + {\rm H.c.},
\end{equation}
and $\tau= \kappa t$ is the effective interaction time (${\rm
H.c.}$; Hermitian conjugate) \cite{Kok}, and
\begin{equation}
\label{SPDC1} |\psi^{d}_{-} \rangle_{\rm AB} = {1 \over \sqrt{d}}
\sum^{d-1}_{i=0} (-1)^i |d-i-1,i \rangle_A |i,d-i-1\rangle_B ,
\end{equation}
where $|\psi^{1}_{-} \rangle_{AB}\equiv |\psi^{1} \rangle_{AB}$ is
a vacuum state in modes $A$ and $B$. For simplicity, we only
consider the first three terms in Eq.~(\ref{SPDCphotons}), and the
normalized state is
\begin{equation}
\label{SimSPDC01} |\Psi \rangle_{AB} = {1 \over \sqrt{N}} \left(
\alpha_1 | \psi^{1}_{-} \rangle_{A B} + \sqrt{2} \alpha_2 |
\psi^{2}_{-} \rangle_{A B} + \sqrt{3} \alpha_3 | \psi^{3}_{-}
\rangle_{A B} \right),
\end{equation}
where
\begin{equation}
N = (\alpha_1)^2 + 2(\alpha_2)^2 + 3 (\alpha_3)^2,
\end{equation}
and
\begin{equation}
\alpha_d \equiv \tanh^{d-1}\tau / \cosh^{2}\tau.
\end{equation}
When a polarization rotator $U(\pi/2)$ is applied in modes $B$
with angle rotation $\pi/2$ (see \ref{Append01}), the state is
equal to
\begin{equation}
\label{SPDC2} |\Psi' \rangle_{AB} = \left[ \hat{I}_A \otimes U_B
(\pi/2) \right] |\Psi \rangle_{AB},
\end{equation}
\begin{equation}
= {1 \over \sqrt{N}} \left( \alpha_1 | \psi^{1} \rangle_{A B} -
\sqrt{2} \alpha_2 | \psi^{2} \rangle_{A B} + \sqrt{3} \alpha_3 |
\psi^{3} \rangle_{A B} \right),
\end{equation}
where $\hat{I}$ is the identity operator. The success probability
to obtain the qutrit state $| \psi^{3} \rangle_{A B}$ is ${3
\alpha_3^2 / N}$.  Experimentally \cite{Howell}, the detection of
$| \psi^{3} \rangle_{A B}$ is performed post-selectively by
sending the two photons through a PBS followed by two number
resolving detections (themselves possible cascaded threshold
detectors \cite{Barry}).

\section{Building a heralded two-qutrit entangled state}
\label{sec3} As PDC is a probabilistic emission process including
vacuum and higher-order photons, a problem arises if we wish to
use these photons in scalable approaches to practical quantum
information processing. Since the photons are not heralded, all
input photons must be measured to ensure that desired states are
prepared - these are often used the schemes of post-selection
\cite{Again}. We now consider an alternative to generate a
heralded qutrit state encoded in the photon polarization (no
longer a vacuum state and other higher-order photons). The qutrit
state is constructed from two heralded Bell pairs and conditioned
on a null detection performed by two ideal threshold detectors.

\begin{figure}[t]
\centering \hspace{-3.5cm}
\includegraphics[width= 9cm,angle=-90]{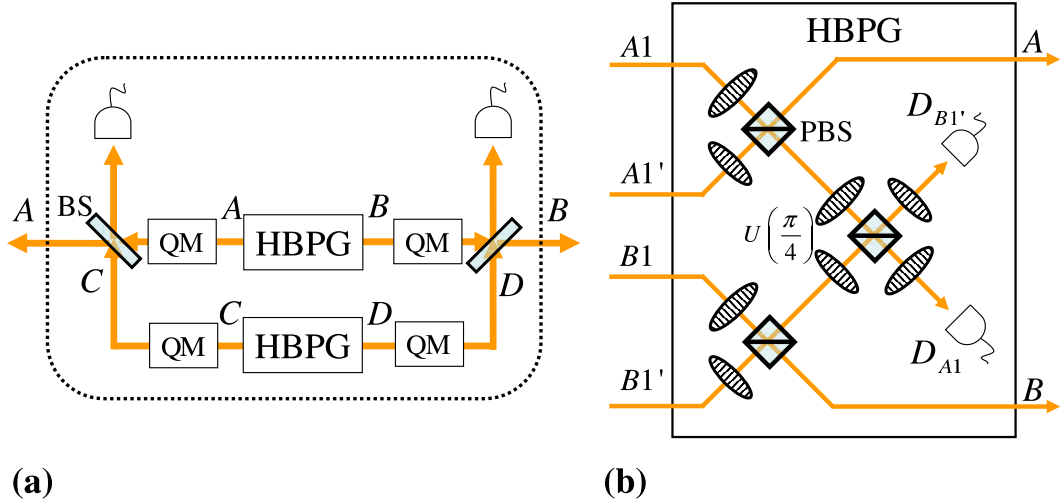}
\vspace{-1cm} \caption{ \label{fig3} (a) A setup is depicted to
build a heralded entangled two-qutrit state. This setup consists
of two BSs, two ideal threshold detectors, two heralded Bell-pair
generators (HBPGs) and four quantum memories (QMs). (b) A circuit
of HBPG consists of three PBSs, eight polarization rotators, and
two detectors \cite{Terry08}. If two out of four single photons
are simultaneously detected in detectors $D_{A1}$ and $D_{B1'}$
respectively, the outcome state is a Bell pair in modes $A$ and
$B$. (BS; beam-splitter, PBS; polarizing beam-splitter,
$U(\pi/4)$; polarization rotator with angle $\pi/4$, and $D_{K}$;
detector in mode $K$)}
\end{figure}

\subsection{Building two heralded Bell pairs}

Let us first assume that heralded Bell-pair generators (HBPGs)
and quantum memories are prepared as shown in Fig.~\ref{fig3}(a).
The two heralded Bell pairs are emitted in modes $A$ and $B$, and
$C$ and $D$. The HBPG has been studied both theoretically and
experimentally in various physical systems \cite{QD1,HeraldEBP}.
In fact an experimental realization has recently been achieved
via a bi-exciton cascade in semiconductor quantum dots
\cite{QD1}, while the theoretical approach of
Ref.~\cite{HeraldEBP} described how to probabilistically produce
heralded two-photon entangled states using controlled-NOT
operations on two PDC sources. Note that even if a heralded
two-photon source is available, four quantum memories (e.g.,
atomic ensembles, cavity QED systems, optical loops \cite{Choi})
will be required to ensure that the two heralded Bell pairs
simultaneously arrive at the two BSs of Fig.~\ref{fig3}(a).

Instead of the aforementioned HBPG schemes, heralded single
photon sources can be used to generate heralded Bell pairs, as
depicted in Fig.~\ref{fig3}(b). This circuit creates a heralded
Bell pair in modes $A$ and $B$ by destroying two out of four
input single photons \cite{Terry08}. The photons enter with
horizontal polarization and are rotated by the polarization
rotator $U(\pi/4)$ that makes them an equal superposition of
horizontal and vertical polarizations. The state emerging from
each of the PBS is such that it is a superposition of the Bell
states (see Eq.~(\ref{Bellpair}))
\begin{equation} \hspace{-1cm}
(|1,0\rangle_{A1}|1,0\rangle_{A1'} +
|0,1\rangle_{A1}|0,1\rangle_{A1'}) \otimes
(|1,0\rangle_{B1}|1,0\rangle_{B1'} +
|0,1\rangle_{B1}|0,1\rangle_{B1'}),
\end{equation}
and other terms that have the two photons in the same spatial
mode, which cannot lead to acceptable detections in the end. A
Type-II fusion gate in modes $A1$ and $B1'$ then fuses the two
Bell pairs from the PBS. If the two single-photon detectors
($D_{A1}$ and $D_{B1'}$) are simultaneously triggered, the output
state is equal to a Bell pair in modes $A$ and $B$. The optimal
success probability to obtain a heralded Bell pair is 1/4 by this
method. \cite{Terry08}.

Any one of several proposed schemes for generating heralded
single photons can be used \cite{HeraldedSP} in the circuit of
Fig.1(b). A simple method is to use PDC photons. Naively one may
think we need to use 8 PDC pairs, and detection of the 4 idler
photons, to generate the four required single photons. In fact,
only two PDC events suffice to create a heralded Bell pair. The
key idea is that a single higher order PDC emission can generate
two single photons simultaneously, as can be seen by studying
Eq.~(\ref{SimSPDC01})) and noting the $|\psi_3^-\>$ term in the
wavefunction. Looking at this term we see that if one horizontal
and one vertical photon are detected in mode $B$, the photons in
mode $A$ are in the state such as $|1,1\rangle_{A}$. That is, we
have picked out the $|\psi_3^-\>$ part of the wavefunction. These
two photons can be easily used as two of the desired single
photons. A separate event from a different PDC would be required
to generate the other two single photons. This process is quite
feasible - a similar idea was used to demonstrate a
path-entangled state in Ref.~\cite{Bouw}.

\subsection{Heralded two-qutrit state from two heralded Bell pairs}

A priori one might try and use two pairs of PDC photons in place of the two
heralded Bell pairs that we described above. That is, let us
assume that two PDC photon pairs (one input in modes $A$ and $B$,
and the other in $C$ and $D$) are merged into two BSs in
Fig.~\ref{fig3}(a), and two null detections are successfully
achieved. In fact this direct approach produces \emph{no} effective
gain in the output state, (over the qutrit state which appears as a higher order effect in the PDC anyway) which means that the output photons are
in exactly the same state as the input state. Thus, we focus here on
the case that two heralded Bell pairs ($|\psi^{2} \rangle_{\rm AB}$ and
$|\psi^{2} \rangle_{CD}$) are simultaneously prepared and
sent to two BSs with the help of quantum memories, as shown in
Fig.~\ref{fig3}(a). The initial state is given by
\begin{equation}
 |\psi^{2}\rangle_{ AB} \otimes |\psi^{2}\rangle_{ CD}.
\end{equation}
A 50:50 BS mixes modes $A$ and $C$, and the other BS does $B$ and
$D$ (see \ref{Append01}).
\begin{equation}
({\rm BS}_{AC} \otimes {\rm BS}_{BD} ) |\psi^{2}\rangle_{ AB}
\otimes |\psi^{2}\rangle_{ CD}.
\end{equation}
The output state $|\Psi^{\rm out}\rangle_{ABCD}$ is
\begin{equation}
|\Psi^{\rm out}\rangle_{ABCD} = - {\sqrt{3}\over 4} |\psi^{3}
\rangle_{\rm AB}\otimes|0,0\rangle_{C} \otimes|0,0\rangle_{D}
        + |\psi^{\rm rej}\rangle_{ABCD}
\end{equation}
where the rejected state $|\psi^{\rm rej}\rangle_{ABCD}$ has one
or more photons in either mode $C$ or $D$.

At this stage, an ideal threshold detector can be used for a
perfect null detection \cite{Barry}. The projection operator
valued measures (POVM) for the ideal threshold detector in mode
$C$ is given by
\begin{equation}
\label{threshold} \{ \Pi_0 = |0,0\rangle_C \langle 0,0|,~~~
\Pi_{>0} = \hat{I} - |0,0\rangle_C \langle 0,0| \}.
\end{equation}
After two null detections in modes $C$ and $D$, the final state
is equal to
\begin{equation}
\left( \hat{I}_{A} \otimes \hat{I}_B \otimes
|0,0\rangle_{C}\langle 0,0| \otimes|0,0\rangle_{D}\langle
0,0|\right) |\Psi^{\rm out}\rangle_{ABCD}.
\end{equation}
For this case, all four incoming photons are merged into modes $A$
and $B$. Thus, the heralded two-qutrit state with four photons
$|\psi^{3} \rangle_{\rm AB}$ can be generated with success
probability $3/16$.

\begin{figure}[t]
\centering
\includegraphics[width= 13cm]{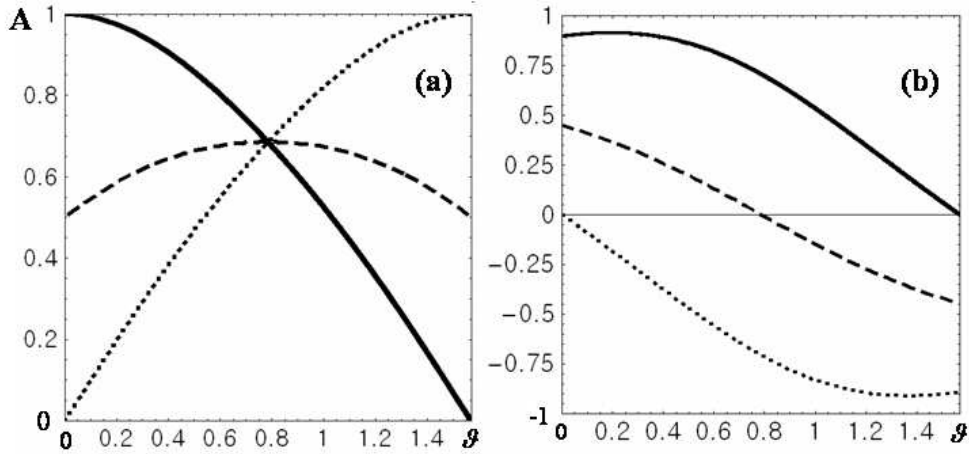}
\vspace{-2cm} \caption{ \label{fig2} Normalized amplitudes (A) of
each terms in Eq.~(\ref{eq:general2Qu}) are depicted for (a)
$\varphi=0$ and (b) $\varphi=\pi$. The solid (dotted) line denotes
$\cos \vartheta / \sqrt{N'}$ (${\rm e}^{{\rm i} \varphi} \sin
\vartheta / \sqrt{N'}$), and the dashed line does $(\cos
\vartheta + \sin \vartheta) / 2\sqrt{N'}$ for (a) and $(\cos
\vartheta - \sin \vartheta) / 2\sqrt{N'}$ for (b).}
\end{figure}

In addition, we are able to produce an unbalanced two-qutrit
entangled state through the same circuit. When an unbalanced Bell
state is prepared in modes $A$ and $B$ (instead of $|\psi^{2}
\rangle_{\rm AB}$) such as
\begin{equation}
|\tilde{\psi}^{2} \rangle_{AB} = \cos \vartheta | 1 , 0 \rangle_A
| 1 , 0  \rangle_B + {\rm e}^{{\rm i} \varphi} \sin \vartheta | 0
, 1  \rangle_A | 0 , 1  \rangle_B,
\end{equation}
the two BSs mix this state with the balanced qutrit state
$|\psi^{2} \rangle_{CD}$. With the null detections in modes $C$
and $D$, the output state becomes
\begin{equation}
\label{eq:general2Qu} \hspace{-2.1cm} |\tilde{\psi}^{3} \rangle =
{\cos \vartheta  \over \sqrt{N'}} | 2,0 \rangle_A | 2,0 \rangle_B
+ {\cos \vartheta + {\rm e}^{{\rm i} \varphi} \sin \vartheta
\over 2 \sqrt{N'}} | 1,1 \rangle_A | 1,1 \rangle_B + {{\rm
e}^{{\rm i} \varphi} \sin \vartheta  \over \sqrt{N'}} | 0,2
\rangle_A | 0,2 \rangle_B ,
\end{equation}
where $N'$ is a normalization factor. In Fig.~\ref{fig2}, the
amplitudes of each terms in Eq.~(\ref{eq:general2Qu}) are shown
for $\varphi=0$ and $\varphi=\pi$ with respect to angle
$\vartheta$. One sees that three amplitudes are the same with
$\varphi=0$ and $\vartheta=\pi/4$ in Fig.~\ref{fig2}(a), and this
maximally entangled qutrit state is the same as
Eq.~(\ref{eq:newstate03}). If $\varphi=\pi$ and $\vartheta=\pi/4$
in Fig.~\ref{fig2}(b), the middle term in
Eq.~(\ref{eq:general2Qu}) disappears and the state equals to
\begin{equation}
\label{FourBell} {1\over \sqrt{2}} (|2,0\rangle_A |2,0\rangle_B -
|0,2\rangle_A |0,2\rangle_B ),
\end{equation}
that is equivalent to a Bell pair with four photons.

\section{Applications}
\label{sec4} A variety of types of generating photonic qutrit
entanglement and testing it by Bell inequality \cite{Bell}
violations have been proposed or implemented
\cite{Aspect,Drummond}. For testing the CGLMP inequality in
particular, two such schemes are using orbital angular momentum
\cite{Vaziri02} or energy-time \cite{Gisin04} entanglement have
been experimentally demonstrated. On the theory side a
two-photon, two-qutrit state (using multi-modes for the higher
dimensions rather than multiple photons) was studied in
\cite{advantage,Zeilinger97}. In Ref.~\cite{Howell}, a PDC photon
pair (including higher-order photon emissions) was used to test a
type of Clauser-Horne-Shimony-Holt (CHSH) inequality proposed by
Gisin and Peres \cite{Bellinequal01}. These schemes all used
coincident detection (post-selection). Here we will show how to
use our scheme to demonstrate a violation of the CGLMP inequality
known as a ``tight Bell inequality'' for a bipartite qudit state
\cite{Masanes}. At the end of this section, we suggest two other
possible applications of the heralded qutrit state. The state
could be compatible with a scheme for secure quantum
communication proposed very recently in Ref.~\cite{Eisenberg}. A
non-maximally entangled two-qutrit state could be use of
conclusive teleportation probabilistically. We also show how a
heralded two-qudit entangled state can be generated by a nested
approach of our qutrit scheme.

\subsection{CGLMP inequality test}
First, the heralded qutrit state is applicable to test the CGLMP
inequality \cite{Bellinequal02}. This inequality is a generalized
form of the CHSH inequality \cite{CHSH} for a bipartite qudit
state, and all the entangled states violate the CGLMP inequality
for qudits \cite{Chen2}. The scenario for the CGLMP inequality
test involves two parties ($A$ and $B$). Party $A$ can perform
one of two possible measurements ($A_1$ or $A_2$) and likewise
party $B$ can perform $B_1$ or $B_2$ just as for the standard CHSH
inequality test. For qutrits, each measurement has three possible
outcomes such as
\begin{equation}
A_1,A_2,B_1,B_2 \in {0,1,2}.
\end{equation}
The CGLMP inequality consists of four functions of the joint
probabilities of the observed outcomes, and each function can
take values $\pm 1$ in general (details are given in Section
4.3). However, for example, if three of them are equal to $+1$,
the other should be $-1$ due to the constraint described by local
realistic theories. The form of the CGLMP inequality is to show
that this particular sum of correlations cannot exceed 2
\cite{Bellinequal02} in a local theory, whereas a value of 2.8729
is achievable by quantum mechanical correlations. This maximum
value cannot be achieved using photons in our scheme, because we
restrict ourselves to using only linear optical elements for
performing the different measurement settings. As such we find
the maximum value our scheme can achieve is $2.5295$. Perfect
photon counting detectors and no photon loss are assumed here.
The photon counters can be constructed as an array of the ideal
threshold detectors as in Eq.~(\ref{threshold}) \cite{Barry}.

\subsection{local operation on a single qutrit}
\begin{figure}[t]
\centering \hspace{-2cm}
\includegraphics[width= 8cm,angle=-90]{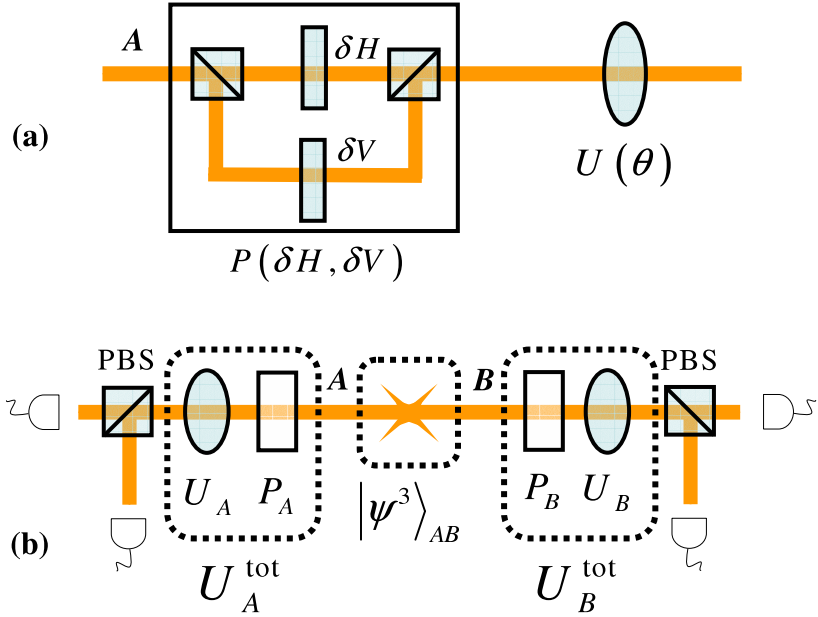}
\vspace{1cm} \caption{ \label{setup01} (a) A total single-qutrit
operation $U^{\rm tot}_{\rm A} (\theta,\delta H, \delta V)$ is
decomposed by two operations ($P(\delta H,\delta V)$ and
$U(\theta)$) in mode $A$. Two phase shifters with different
angles are located between two PBSs to perform the operation
$P(\delta H,\delta V)$. (b) A setup for CGLMP inequality test is
depicted.}
\end{figure}

Let us consider first what kind of single-qutrit operations are
available on the bunched photons. Unfortunately, it is impossible
to perform an arbitrary single-qutrit operation (even a quantum
Fourier transform say) because of the restricted bi-photon
operation. The best we can do is a U(2) operation on each photon
in the same spatial mode, and this makes a certain single-qutrit
operation defined by $U^{\rm tot}$. As shown in
Fig.~\ref{setup01}(a), the operator ${U}_3 (\theta)$ is built by a
polarization rotator and ${P}_3 (\delta H, \delta V)$ does using
two different phase shifters between the two PBSs (see
\ref{Append01}). Thus, we here use a simple realization of a
single-qutrit operation represented by
\begin{equation}
U^{\rm tot} (\theta, \delta H, \delta V) = {U}_{3}\, (\theta)~
{P}_{3}\, (\delta H, \delta V).
\end{equation}

\subsection{CGLMP inequality and four
functions of the joint probabilities}

We now explain how to test the CGLMP inequality
\cite{Bellinequal02} on the two-qutrit state in
Eq.~(\ref{eq:newstate03}). According to Ref.
\cite{Bellinequal02}, the value of the CGLMP inequality for
qutrits is given by
\begin{equation}
\label{eq:3inequal01} I_3 = \sum^{4}_{i=1} {\cal B}_i \le2,~~~
\end{equation}
where four functions are represented by
\begin{equation}
{\cal B}_1 =P(A_1=B_1) - P(A_1=B_1-1),
\end{equation}
\begin{equation}
{\cal B}_2= P(B_1=A_2+1) -P(B_1=A_2),
\end{equation}
\begin{equation}
{\cal B}_3 = P(A_2=B_2) - P(A_2=B_2-1),
\end{equation}
\begin{equation}
{\cal B}_4 = P(B_2=A_1) - P(B_2=A_1-1),
\end{equation}
with a sum of joint probabilities
\begin{equation} P(A_i=B_j + k)=\sum^{2}_{l=0} P(A_i=l,B_j=k+l ~
{\rm mod}~3).
\end{equation}
In order to violate the condition $I_3 \le 2$, we need to choose
suitable sets of local operations on both sides ($U^{\rm
tot}_{A_1}$, $U^{\rm tot}_{A_2}$, $U^{\rm tot}_{B_1}$, and $U^{\rm
tot}_{B_2}$) and calculate joint probabilities by measuring the
qutrit state in modes $A$ and $B$ individually.

To understand what the qutrit operations do, it is easiest to
check what is the value for each term ${\cal B}_i$ in
Eq.~(\ref{eq:3inequal01}). As an example, let us choose two local
setups $A_1$ and $B_1$ such that $U^{\rm tot}_{A_1} (\theta,0, 0)$
and $U^{\rm tot}_{B_1}(\theta,\delta H, 0)$. To obtain the first
term of joint probabilities
\begin{equation}
    {\cal B}_1= {}_{\rm AB} \langle {\Psi}^{11} | {O} |{\Psi}^{11}
        \rangle_{\rm AB},
\end{equation}
the observable is given by
\begin{equation} \hspace{-1cm}
O = \left[\sum_{i=0}^{2} |i\rangle_{\rm A}\langle i| \otimes
|i\rangle_{\rm B} \langle i|\right] - \left[ \sum_{i=0}^{2}
|i\rangle_{\rm A}\langle i| \otimes |i+1~{\rm mod}~3\rangle_{\rm
B} \langle i+1~{\rm mod}~3|\right],
\end{equation}
and the state after the operations is equal to
\begin{equation}
|\Psi^{11} \rangle_{\rm AB} = \left[ U^{\rm tot}_{A_1} (\theta,0,
0) \otimes U^{\rm tot}_{B_1}(\theta,\delta H, 0) \right] |\psi^{3}
\rangle_{\rm AB}.
\end{equation}
The function of joint probabilities ${\cal B}_1$ is limited by
\begin{equation}
-{1\over3} \le {\cal B}_1 \le 1
\end{equation}
and this result shows that the minimum value of ${\cal B}_i$ is
$-{1/3}$ \cite{arccos}. Without the restriction of linear optical
elements ${\cal B}_i$ can approach to -1.

\begin{figure}[t]
\centering
\includegraphics[width= 7.5cm,angle=-90]{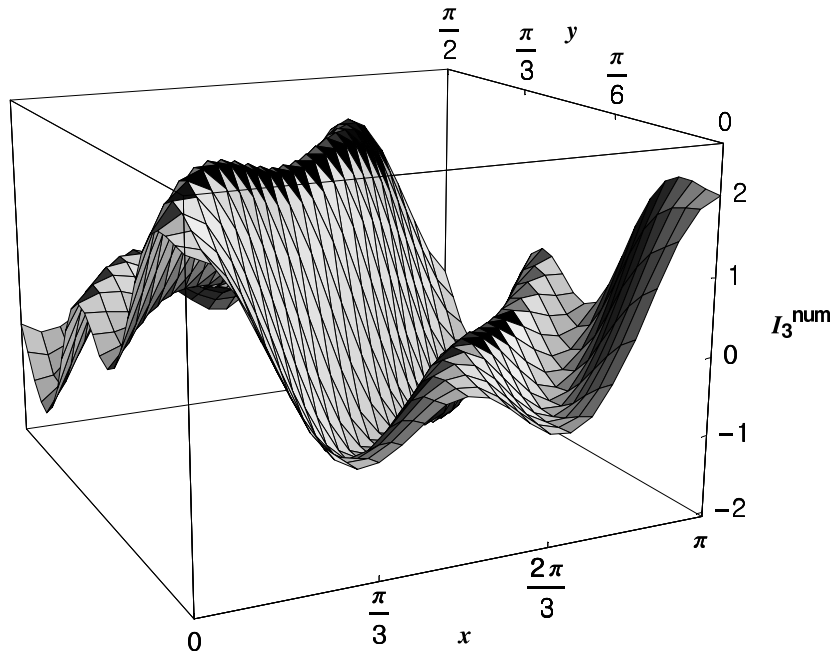}
\vspace{1cm}
 \caption{ \label{fig5} $I^{\rm num}_3$ in Eq.~(\ref{eq:3inequal01})
with $U^{\rm tot}_{A_1} ({\pi/ 4},x,y)$, $U^{\rm tot}_{B_1} ({\pi/
4},x,2y)$ , $U^{\rm tot}_{A_2} ({\pi/ 4},x,3y)$, and $U^{\rm
tot}_{B_2} ({\pi/4},x,0)$. A maximum value is approximately
$2.5295$ with $x = y \approx 0.4507$ and $\theta =\pi/4$.}
\end{figure}

As a numerical result, the maximum value of $I_3$ in
Eq.~(\ref{eq:3inequal01}) ideally reaches $4/(6\sqrt{3}-9)\approx
2.87293$ \cite{Bellinequal02}. To see how close we can come to
this with our restricted operations we numerically examine its
violation with 12 parameters in four sets of unitary operators
such that $U^{\rm tot}_{A_1} (\theta_{A_1},\delta H_{A_1} ,\delta
V_{A_1})$, $U^{\rm tot}_{B_1} (\theta_{B_1},\delta H_{B_1} ,\delta
V_{B_1})$ , $U^{\rm tot}_{A_2} (\theta_{A_2},\delta H_{A_2}
,\delta V_{A_2})$, and $U^{\rm tot}_{B_2} (\theta_{B_2},\delta
H_{B_2} ,\delta V_{B_2})$. In Fig.~\ref{fig5}, the value $I_3$ is
depicted with two variables ($x$, $y$) for fixed $\theta=\pi/4$.
Its maximum value reaches to $2.5295$, and this clearly shows
that the maximally entangled qutrit state can violate the CGLMP
inequality under the restrictions considered.

\subsection{Remarks regarding the inequality test}

Space-like separation between the system
creating the qutrit state and the measurements performing the  Bell
inequality tests on the two-qutrit state  would be required. That is,
 the detection of the photons should be undertaken
with the null detections done early enough before testing the
qutrit state for violating Bell's inequality.

In the quantum channel of our model, we assumed a perfect channel
(no photon loss).  We note that in the case of PDC the entangled photon pairs are tightly
correlated between the frequency and time domains
\cite{Rohde}. Because this results in undesired
distinguishability of the incoming photons, spectral filters must
be positioned in front of detectors. This reduces the rate of
successful events and can also be treated as an additional
contribution to a finite detection efficiency.

We have also assumed a perfect photon counter is used. Such can
be effectively constructed by an array of ideal threshold
detectors \cite{Barry}. Note that imperfect sources and finite
detection efficiency would open the same loophole problems for
the CGLMP inequality test as they do for the CHSH test. As with
those tests, the ``no-enhancement'' or ``fair sampling''
assumption would need to be invoked to argue the implausibility
of these loopholes. This assumption is basically that the
purported hidden variables cannot exploit the finite efficiencies
of the sources and detectors by changing themselves from run to
run of the test in a way which controls whether they will be
detected or not. Alternatively, the Clauser-Horne (CH) inequality
\cite{CH} can be used when only imperfect sources and detectors
are available. This inequality is more robust because
``null-detections'' (due either to an inefficient detector or a
probabilistic source) are taken into account as actual events
which enter the correlation functions from which the CH
inequality is constructed. The end result is that this inequality
is more robust to such noise, and requires less assumptions about
fair-sampling. A similar test has been studied for a (different)
qutrit system in Ref.~\cite{Kwek}.

\subsection{Other possible applications}

The heralded qutrit state has a potential to utilize for quantum
cryptography. Very recently, several interesting protocols have
been investigated in qutrits for secure quantum key distribution
and shown that the qutrit cyptosystem is more secure than
qubit-based protocols against some attacks \cite{Eisenberg}. They
are constructed in the spirit of the Bennett and Brassard scheme
\cite{BB84}. In one of the protocols (called three-ray), nine
basis vectors (grouped in three basis sets) are randomly chosen
by Alice, and a quantum state represented by that vector is sent
to Bob. Once Bob received the state and measured it in a certain
basis set, he publicly announces which basis set he chose.
Finally, Alice confirms whether his choice is correct or not, and
both Alice and Bob share a string of raw trits.

One may consider a protocol of quantum key distribution using the
entangled two-qutrit state $|\psi^3\rangle_{AB}$. In our system,
the three basis sets can be realized by local qutrit operations.
Its security could be checked along the same lines as the Ekert
protocol in qubits \cite{Ekert}. In analogy with Ekert protocol,
once the choice of local operations are matched in both parties,
it becomes a raw trit key otherwise they could test a violation of
a certain inequality. However, it is an open question which
inequality can provide a violation of local hidden theory for
this system even though many Bell's inequalities have been
studied theoretically for a bipartite system
\cite{Chen,Bellinequal03}. Therefore, it will be worth studying
that which inequality for qutrits could be violated by the qutrit
setups.

A two-qutrit state can be used to perform quantum teleportation
probabilistically \cite{ConcT}. As an example, an arbitrary
single-qutrit state is prepared in mode $C$ proposed in
Ref.~\cite{Li}. Using the circuit shown in Fig.~\ref{fig3}(a),
the output state in modes $A$ and $B$ can be made as an
unbalanced two-qutrit state (see \ref{Append02})
\begin{equation}
\label{unbal01} {1\over3} (2| 0 \rangle_A | 0 \rangle_B + | 1
\rangle_A | 1 \rangle_B + 2| 2 \rangle_A | 2 \rangle_B ).
\end{equation}
After a PBS between modes $B$ and $C$ and a polarization rotator
$U(\pi/4)$ in both $B$ and $C$, the unknown state in mode $C$ can
be teleported to mode $A$ conclusively by destroying four photons
in modes $B$ and $C$. Four successful detections indicate that
the conclusive teleportation is achieved, and the total success
probability is 1/9.

The other possible application is that a heralded two-qudit state
can be generated by a nested approach of Fig.~\ref{fig3}(a). Let
us consider that state $|\psi^{d-1} \rangle_{AB}$ is prepared in
Fig.~\ref{fig3}(a) instead of $|\psi^{2} \rangle_{AB}$. After two
BSs, the case of the null detections on modes $C$ and $D$
repeatedly, the state $|\psi^{d} \rangle_{AB}$ in
Eq.{\ref{eq:photonDstate01}) is achieved with a success
probability
\begin{equation}
\label{eq:successP01} P_{d} = {d(d-1) \over 2^{2d-1}}.
\end{equation}
This circuit can be iteratively used to generate a maximally
entangled qudit state. Thus, a heralded two-qudit entangled state
can be made from $d-1$ heralded Bell pairs using $2(d-2)$ ideal
threshold detectors.

\section{Conclusion}
\label{sec5}

We have analyzed a simple scheme for generation of a heralded
two-qutrit entangled state in polarized photons. To achieve the
qutrit state, when two heralded Bell pairs are mixed by two BSs,
two null detections guarantee the generation of the heralded
two-qutrit state. We have analyzed the extent to which the qutrit
state can be used to demonstrate violation of the CGLMP
inequality when measurements are restricted to those achievable
with linear optics and photo-detection. We discussed how a source
of such states could be compatible with a cryptography scheme
proposed in Ref.~\cite{Eisenberg}. We also suggested two possible
methods for teleporting an unknown qutrit and creating a heralded
two-qudit entangled state by a nested version of our scheme.

\section*{Acknowledgements}
JJ thanks Jeong San Kim, Anthony Laing, Jeremy O'Brien, Y.-C.
Liang, and Artur Scherer for useful discussions. BCS appreciates
financial support from iCORE, MITACS, QuantumWorks, and a CIFAR
Associateship. JJ acknowledges PhDPlus, iCORE, MITACS, and NSERC.
TR is supported in part by the United Kingdom Engineering and
Physical Sciences Research Council and the US Army Research
Office.

\appendix

\section{Basic optical tools}
\label{Append01} A 50:50 BS between modes $A$ and $C$ for a
single-photon inputt is given by
\begin{equation}
\label{eq:U03} {\rm BS}_{AC} = {\rm e}^{ {\rm i} {\pi\over 4}
\hat{\cal J}_{BS}},
\end{equation}
where
\begin{equation}
    \hat{\cal J}_{BS} =
        a_H c^{\dagger}_H + a^{\dagger}_H c_H +
        a_V c^{\dagger}_V + a^{\dagger}_V c_V ,
\end{equation}
The transformation of BS between modes $A$ and $C$ for a
single-photon input is given by
\begin{equation}
\label{eq:BS01} |1,0\rangle_{\rm A} \rightarrow { |1,0\rangle_{\rm
A} +  {\rm i} |1;0\rangle_{C} \over \sqrt{2}},~~~{\rm and}~~~
|1,0\rangle_{C} \rightarrow { {\rm i} |1,0\rangle_{\rm A} +
|1,0\rangle_{C} \over \sqrt{2}},
\end{equation}
where state $|1,0\rangle$ can be replaced to $|0,1\rangle$ for
vertically polarized photons.

A polarization rotator with angle $\theta$ is given by
\begin{equation}
\label{eq:U03} {U} (\theta) = {\rm e}^{\theta \hat{\cal
J}_{R}},\end{equation} where
\begin{equation}
    \hat{\cal J}_{R} = a^{\dagger}_V a_H - a^{\dagger}_H a_V .
\end{equation}
For example, if we consider $\left[ \hat{I}_A \otimes U_B (\theta)
\right] |\Psi \rangle_{AB}$, the unitary operator with angle
$\theta$ in mode $B$ is represented by a $6\times 6$ matrix form
\begin{equation}
\label{eq:4Utotal} U_B (\theta) = \left(
\begin{array}{ccc}
1 & 0 & 0  \\
0& U_{2}&  0 \\
0& 0 & U_{3} \\
\end{array} \right),
\end{equation}
where
\begin{equation}
\label{eq:Uqubit} U_{2} = \left(
\begin{array}{cc}
\cos \theta & \sin\theta \\
- \sin\theta & \cos\theta \\
\end{array} \right),
\end{equation}
\begin{equation}
\label{eq:3U04} U_{3} = \left(
\begin{array}{ccc}
\cos^2 \theta & \sqrt{2} \cos \theta \sin \theta
& \sin^2 \theta \\
- \sqrt{2} \cos \theta \sin \theta & \cos^2 \theta - \sin^2 \theta
&  \sqrt{2} \cos \theta \sin \theta \\
 \sin^2 \theta & - \sqrt{2} \cos \theta \sin \theta & \cos^2 \theta \\
\end{array} \right).
\end{equation}
where the basis vectors of Eq.~(\ref{eq:4Utotal}) are given by
$|0,0\rangle$, $|1,0\rangle$, $|0,1\rangle$, $|2,0\rangle$,
$|1,1\rangle$, and $|0,2\rangle$. A phase shifter in both
polarizations is given by
\begin{equation}
{P} (\delta H, \delta V) = {\rm e}^{{\rm i} \delta H \hat{n}_H}
{\rm e}^{{\rm i} \delta V \hat{n}_V},
\end{equation}
where $\hat{n}_{H}=a^{\dagger}_H a_H$, and
$\hat{n}_{V}=a^{\dagger}_V a_V$ \cite{Nielsen,Barry01}. For
example, the phase operator is mode $A$ given by
\begin{equation}
\label{eq:4Utotal} P_A (\delta H, \delta V) = \left(
\begin{array}{ccc}
1 & 0 & 0  \\
0& P_{2}&  0 \\
0& 0 & P_{3} \\
\end{array} \right),
\end{equation}
where
\begin{equation}
\label{eq:Uqubit} P_{2} = \left(
\begin{array}{cc}
{\rm e}^{{\rm i} \delta H} & 0 \\
0 & {\rm e}^{{\rm i} \delta V} \\
\end{array} \right),
\end{equation}
\begin{equation}
\label{eq:3U04} P_{3} = \left(
\begin{array}{ccc}
{\rm e}^{{\rm i} 2 \delta H} & 0 & 0  \\
0 & {\rm e}^{{\rm i} (\delta H  + \delta V)}
& 0 \\
0 & 0 & {\rm e}^{{\rm i} 2 \delta V}  \\
\end{array} \right),
\end{equation}
where the basis vectors are the same as Eq.~(\ref{eq:4Utotal}).

\section{Building an unbalanced two-qutrit state}
\label{Append02} We start with the state given by
Eq.~(\ref{FourBell}) in modes $A$ and $B$. When a relative phase
is changed by two PBSs and a polarization rotator $U(\pi/2)$, the
state becomes
\begin{equation}
\label{FourBell2} |S1\rangle_{AB} = {1\over \sqrt{2}}
(|2,0\rangle_A |2,0\rangle_B + |0,2\rangle_A |0,2\rangle_B ).
\end{equation}
To achieve the unbalanced state in Eq.~(\ref{unbal01}), an
additional entangled state is required in modes $C$ and $D$ such
as
\begin{equation}
\label{unBell} |S2\rangle_{CD} = {1\over 17} (4\sqrt{17}
|1,0\rangle_C |1,0\rangle_D + \sqrt{17} |0,1\rangle_C
|0,1\rangle_D ).
\end{equation}
When the states in Eq.~(\ref{FourBell2}) and Eq.~(\ref{unBell})
are merged into the circuit shown in Fig.~\ref{fig3}(a), the
output state becomes
\begin{equation}
({\rm BS}_{AC} \otimes {\rm BS}_{BD} ) |S1\rangle_{AB} \otimes
|S2\rangle_{CD}.
\end{equation}
After the detection of a horizontally polarized photon in modes
$C$ and $D$, respectively, the outcome state becomes the state in
Eq.~(\ref{unbal01}).


\begin{thebibliography}{9}
\bibitem{arccos}
$U^{\rm tot}_{A_1} (\pi/4,0,0)$ and $U^{\rm tot}_{B_1} (\pi/4, 2
\arccos({1/ \sqrt{3}}),0)$ makes the minimum value -1/3. The
operations $U^{\rm tot} (\pi/4,0,0)$ and $U^{\rm tot} (\pi/4,
\delta H_0,0)$ are equal to
\begin{equation}
\label{eq:arc} \hspace{-1cm} {1\over 2} \left(
\begin{array}{ccc}
1 & \sqrt{2} & 1 \\
-\sqrt{2} & 0 & \sqrt{2} \\
1 & -\sqrt{2} & 1 \\
\end{array} \right),
~~~{\rm and}~~~ {1\over 2} \left(
\begin{array}{ccc}
{\rm e}^{{\rm i} 2 \delta H_0} & \sqrt{2} {\rm e}^{{\rm i} \delta H_0} & 1 \\
- \sqrt{2} {\rm e}^{{\rm i} 2 \delta H_0} & 0 & \sqrt{2} \\
{\rm e}^{{\rm i} 2 \delta H_0} & - \sqrt{2} {\rm e}^{{\rm i} 2 \delta H_0} & 1 \\
\end{array} \right),
\end{equation}
respectively, where $\delta H_0$ = 2$\arccos(1/\sqrt{3})\approx
1.91063$.
\end{thebibliography}
\end{document}